\documentstyle[preprint,aps]{revtex}

\draft

\begin{document}
\title{Spin-orbital gapped phase with least symmetry breaking in the
one-dimensional symmetrically coupled spin-orbital model}
\author{Guang-Ming Zhang$^1$, Hui Hu$^2$, and Lu Yu$^3$}
\address{$^1$Center for Advanced Study, Tsinghua University, Beijing 100084, China\\
$^2$The Abdus Salam International Center for Theoretical Physics, P.O. Box\\
586, 34100, Trieste, Italy\\
$^3$Institute of Theoretical Physics and Interdisciplinary Center of\\
Theoretical Studies, Academia Sinica, Beijing 100080, China}
\date{\today}
\maketitle

\begin{abstract}
To describe the spin-orbital energy gap formation in the one-dimensional
symmetrically coupled spin-orbital model, we propose a simple mean field
theory based on an SU(4) constraint fermion representation of spins and
orbitals. A spin-orbital gapped phase is formed due to a marginally relevant
spin-orbital valence bond pairing interaction. The energy gap of the spin
and orbital excitations grows extremely slowly from the SU(4) symmetric
point up to a maximum value and then decreases rapidly. By calculating the
spin, orbital, and spin-orbital tensor static susceptibilities at zero
temperature, we find a crossover from coherent to incoherent magnetic
excitations as the spin-orbital coupling decreasing from large to small
values.
\end{abstract}


\tighten


It is a very interesting problem to look for exotic quantum magnetic states
realized with the spin and orbital degrees of freedom. Since the discovery
of new quasi-one-dimensional spin gapped materials Na$_2$Ti$_2$Sb$_2$O \cite
{axtell} and NaV$_2$O$_5$ \cite{isobe}, there has been considerable interest
in magnetic systems with orbital degeneracy \cite
{pati,yqli,mila,azaria,ysu,itoi,zheng,parola}. It is believed that the
unusual magnetic properties observed in these compounds can be explained by
a simple two-band Hubbard model at quarter filling, and in the large Coulomb
repulsion limit the effective Hamiltonian is simplified to a model of two
symmetrically coupled spin -1/2 Heisenberg chains: \cite{kugel,feiner,aa95} 
\begin{eqnarray}
H &=&J\sum_i\left( {\bf S}_i\cdot {\bf S}_{i+1}+{\bf T}_i\cdot {\bf T}%
_{i+1}\right)  \nonumber \\
&&+V\sum_i({\bf S}_i\cdot {\bf S}_{i+1})({\bf T}_i\cdot {\bf T}_{i+1}),
\end{eqnarray}
where ${\bf S}_i$ and ${\bf T}_i$ denote the spin and orbital degrees of
freedom, respectively. Here both coupling parameters $J$ and $V$ are assumed
to be antiferromagnetic, and the model Hamiltonian is SU(2)$\otimes $SU(2)
symmetric with an additional Z$_2$ symmetry in exchange between ${\bf S}_i$
and ${\bf T}_i$.

For this model, the critical point $V=0$ describes two independent isotropic
Heisenberg spin-1/2 chains with gapless excitations. In the weak coupling
regime $V/J\ll 1$, it has been shown that the model describes a non-Haldane
spin liquid where magnetic excitations are gapful but {\it incoherent} \cite
{tsvelik}. In the strong coupling regime\cite{convention} $V/J\gg 1$,
however, a special point $V/J=4$ has been identified where the Hamiltonian
becomes SU(4) invariant \cite{aa95}, and it has been demonstrated by Bethe
ansatz and effective field theory methods that the low-energy excitations
are {\it coherent,} given by three branches of gapless elementary
excitations \cite{sutherland,affleck}. However, as emphasized in Ref. \cite
{azaria}, there is no renormalization group flow from the first to the
second critical points due to the Zamolodchikov theorem on the central
charge. In the Heisenberg limit ($V=0$), the total central charge is $c=2$,
while $c=3$ at the SU(4) symmetric point. This means a gapped phase is
expected in-between the weak and strong coupling limits with gapless
magnetic excitations. In particular, it has been shown \cite{km} that when $%
V/J=4/3$, the model has an exact ground state in which spin and orbital
operators may form dimerized singlets in a staggered pattern, and such a
matrix product state is doubly degenerate and gapped. However, so far it is
not clear whether such staggered dimmerized singlets can represent the
ground state in the whole gapped phase. On the other hand, a crossover
transition from incoherent to coherent magnetic excitations is speculated as
the spin-orbital coupling changing from small to large values within the
gapped phase \cite{azaria}. Since the perturbation treatment from either end
can not provide a unified description, a non-perturbative interpolation
scheme would be highly desirable.

The purpose of the present paper is to develop a simple mean field (MF)
theory based on the strong-coupling SU(4) symmetric limit and to describe
such a spin-orbital gap formation along with the coherent-incoherent
crossover of magnetic excitations. When $V/J$ $<4$ and is close to the
strong-coupling SU(4) point, the spin and orbital static susceptibilities
display very sharp coherent magnon peaks at the nesting wave vector $%
2k_F=\pi /2$, corresponding to a commensurate spin-density wave of period
four lattice spacings. Away from the strong coupling symmetric point, a
spin-orbital valence bond (VB) pairing interaction is present, leading to an
energy gap in the quasiparticle excitations. The energy gap initially grows 
{\it extremely slowly} from $V/J=4$ up to a maximum value near $V/J\sim
1.265 $, and then decreases rapidly to a very small value near $V/J\sim
0.462 $. From the calculated spin and orbital static susceptibilities, the
coherent magnetic peaks around $q=\pm \pi /2$ are gradually suppressed and 
{\it slightly} shifted, and their spectral weights are transferred to the
incoherent background around $q=\pm \pi $. Moreover, the present MF theory
also provides the correlation spectra of the spin-orbital tensor operators,
exhibiting further clear evidence of the nontrivial crossover from the weak
to strong coupling limits of the model.

First, up to a constant, the model Hamiltonian can be rewritten as: 
\begin{eqnarray}
H &=&J_c\sum_i\left( 2{\bf S}_i\cdot {\bf S}_{i+1}+\frac 12\right) \left( 2%
{\bf T}_i\cdot {\bf T}_{i+1}+\frac 12\right)  \nonumber \\
&&-J_s\sum_i\left( 2{\bf S}_i\cdot {\bf S}_{i+1}-\frac 12\right) \left( 2%
{\bf T}_i\cdot {\bf T}_{i+1}-\frac 12\right) ,
\end{eqnarray}
where the first part corresponds to an SU(4) spin-orbital symmetric model in
the SU(4) {\it fundamental} representation \cite{yqli,mila,azaria,ysu,parola}%
, while the second part corresponds to a {\it staggered} SU(4) spin-orbital
VB model \cite{santoro,mn,zhang-shen,parola} in which alternating sublattice
sites transform according to the SU(4) fundamental and anti-fundamental
representations, respectively. Here the coupling parameters are regrouped
into $J_c=\left( J/2+V/8\right) $ and $J_s=\left( J/2-V/8\right) $.

In the strong coupling SU(4) symmetric point, $V/J=4$, $J_c=J$ and $J_s=0$.
Our MF theory will take this limit as a starting point, while the weak
coupling limit $V=0$ corresponds to the case of $J_c=J_s=J/2$. In order to
maintain the higher symmetry of the strong coupling limit and to
characterize both spin and orbital degrees of freedom at the same time, an
SU(4) constrained fermion representation is introduced, and its generators
are given by $F_\beta ^\alpha (i)=C_{i,\alpha }^{\dagger }C_{i,\beta }$,
satisfying the SU(4) Lie algebra 
\begin{equation}
\left[ F_\beta ^\alpha (i),F_\nu ^\mu (i)\right] =\delta _{\beta ,\mu }F_\nu
^\alpha (i)-\delta _{\alpha ,\nu }F_\beta ^\mu (i).
\end{equation}
The four states we consider are $|+,+>,|-,+>,|+,->$ and $|-,->,$ where the
first index specifies the spin projection, while the second one is the
orbital projection. It's then obvious that the spin and orbital operators
are expressed in terms of these four-component fermions as: 
\begin{eqnarray}
S_i^{+} &=&C_{i,1}^{\dagger }C_{i,2}+C_{i,3}^{\dagger }C_{i,4},\text{ } 
\nonumber \\
S_i^{-} &=&C_{i,2}^{\dagger }C_{i,1}+C_{i,4}^{\dagger }C_{i,3},  \nonumber \\
S_i^z &=&\frac 12(C_{i,1}^{\dagger }C_{i,1}-C_{i,2}^{\dagger
}C_{i,2}+C_{i,3}^{\dagger }C_{i,3}-C_{i,4}^{\dagger }C_{i,4}); \\
T_i^{+} &=&C_{i,1}^{\dagger }C_{i,3}+C_{i,2}^{\dagger }C_{i,4},  \nonumber \\
T_i^{-} &=&C_{i,3}^{\dagger }C_{i,1}+C_{i,4}^{\dagger }C_{i,2},  \nonumber \\
T_i^z &=&\frac 12(C_{i,1}^{\dagger }C_{i,1}+C_{i,2}^{\dagger
}C_{i,2}-C_{i,3}^{\dagger }C_{i,3}-C_{i,4}^{\dagger }C_{i,4}),
\end{eqnarray}
from which the following commutation relations can be proved 
\begin{eqnarray}
\left[ S_i^{+},S_j^{-}\right]  &=&2S_i^z\delta _{i,j},\quad \left[
S_i^z,S_j^{\pm }\right] =\pm S_i^{\pm }\delta _{i,j},  \nonumber \\
\left[ T_i^{+},T_j^{-}\right]  &=&2T_i^z\delta _{i,j},\quad \left[
T_i^z,T_j^{\pm }\right] =\pm T_i^{\pm }\delta _{i,j},  \nonumber \\
\left[ S_i^\alpha ,T_j^\beta \right]  &=&0,\text{ }\alpha ,\beta =x,y,z.
\end{eqnarray}
It is thus demonstrated that the spin and orbital operators are two
independent degrees of freedom and both of them obey their respective SU(2)
Lie algebra. By imposing a local constraint $\sum_\mu C_{i,\mu }^{\dagger
}C_{i,\mu }=1$, we can further prove that constraints ${\bf S}_i^2={\bf T}%
_i^2=3/4$ are satisfied, corresponding to the spin-1/2 system with two-fold
orbital degeneracy. Under the new representation, the model Hamiltonian is
expressed as a quadratic form in terms of two composite operators 
\begin{equation}
H=-J_c\sum_i:A_i^{\dagger }A_i:-J_s\sum_iB_i^{\dagger }B_i,
\end{equation}
with 
\begin{eqnarray}
A_i &=&\sum_\mu C_{i+1,\mu }^{\dagger }C_{i,\mu },  \nonumber \\
B_i &=&\left[ \left( C_{i+1,4}C_{i,1}+C_{i+1,1}C_{i,4}\right) \right.  
\nonumber \\
&&\left. -\left( C_{i+1,3}C_{i,2}+C_{i+1,2}C_{i,3}\right) \right] ,
\end{eqnarray}
where $A_i$ describes a nearest neighbor VB hopping parameter, while $B_i$
represents a nearest neighbor VB pairing parameter. The normal ordering has
been chosen in the first term.

To develop a MF theory, the nearest neighbor VB order parameters are defined
by $\Delta _c(i)=\langle A_i\rangle $ and $\Delta _s(i)=-\langle B_i\rangle $%
. The model Hamiltonian is then decomposed into 
\begin{eqnarray}
{\cal H} &=&-J_c\sum_{i,\mu }\left[ \Delta _c(i)C_{i,\mu }^{\dagger
}C_{i+1,\mu }+H.c.\right] +\lambda \sum_{i,\mu }C_{i,\mu }^{\dagger
}C_{i,\mu }  \nonumber \\
&&+J_s\sum_i\left[ \Delta _s(i)\left( C_{i,1}^{\dagger }C_{i+1,4}^{\dagger
}+C_{i,4}^{\dagger }C_{i+1,1}^{\dagger }\right) \right.  \nonumber \\
&&\left. -\Delta _s(i)\left( C_{i,2}^{\dagger }C_{i+1,3}^{\dagger
}+C_{i,3}^{\dagger }C_{i+1,2}^{\dagger }\right) +H.c.\right]  \nonumber \\
&&-\lambda N+N\left[ J_c|\Delta _c(i)|^2+J_s|\Delta _s(i)|^2\right] ,
\end{eqnarray}
where a local chemical potential is first introduced to impose the local
constraint and then it is replaced by a global value $\lambda $ keeping the
translational symmetry. When the spatial uniformity of VB parameters are
also assumed, in terms of a generalized Nambu spinor, 
\[
\Psi _k^{\dagger }=\left( C_{k,1}^{\dagger },C_{k,2}^{\dagger
},C_{k,3}^{\dagger },C_{k,4}^{\dagger
},C_{-k,1},C_{-k,2},C_{-k,3},C_{-k,4}\right) 
\]
the MF model Hamiltonian can be rewritten in a compact form 
\[
{\cal H}=\frac 12\sum_k\Psi _k^{\dagger }{\bf H}_{mf}(k)\Psi _k+\lambda
N+N\left( J_c\Delta _c{}^2+J_s\Delta _s{}^2\right) , 
\]
where ${\bf H}_{mf}(k)=\left[ \lambda -\Delta _c(k)\right] {\bf \Omega }%
_1-\Delta _s(k){\bf \Omega }_2$, $\Delta _c(k)=2J_c\Delta _c\cos k$, $\Delta
_s(k)=(2J_s\Delta _s\sin k)$ with ${\bf \Omega }_1=\sigma _z\otimes \sigma
_0\otimes \sigma _0$ and ${\bf \Omega }_2=\sigma _x\otimes \sigma _y\otimes
\sigma _y$. The corresponding Lagrangian is given by 
\[
L_{mf}=\frac 12\sum_k\Psi _k^{\dagger }(i\omega _n)\left[ i\omega _n-{\bf H}%
_{mf}(k)\right] \Psi _k(i\omega _n)+... 
\]
the Matsubara Green's function matrix is thus derived as 
\begin{equation}
{\bf G}(k,i\omega _n)=\frac{i\omega _n+\left[ \lambda -\Delta _c(k)\right]
\Omega _1-\Delta _s(k)\Omega _2}{(i\omega _n)^2-\left[ \lambda -\Delta
_c(k)\right] ^2-\Delta _s^2(k)}.
\end{equation}
Then the fermionic excitation spectra with fourfold degeneracy are yielded 
\begin{equation}
\epsilon _k=\pm \sqrt{\left( \lambda -2J_c\Delta _c\cos k\right)
^2+(2J_s\Delta _s\sin k)^2}.
\end{equation}
In the excitation spectra (with plus sign), the local energy minima appear
at the specific momentum -- the so-called Fermi momentum $k_F$, where an
energy gap opens up when $\Delta _s\neq 0$.

Now consider the static properties at zero temperature. By filling in all
states with negative energies, the ground state energy per site is evaluated
as 
\begin{eqnarray}
\varepsilon _g &=&-\int_{-\pi }^\pi \frac{dk}\pi \sqrt{\left( \lambda
-2J_c\Delta _c\cos k\right) ^2+(2J_s\Delta _s\sin k)^2}  \nonumber \\
&&+\lambda +\left( J_c\Delta _c{}^2+J_s\Delta _s{}^2\right) .
\end{eqnarray}
By minimizing the ground state energy with respect to parameters $\Delta _c$%
, $\Delta _s$, and $\lambda $, the saddle point equations are derived as: 
\begin{eqnarray}
\int_{-\pi }^\pi \frac{dk}\pi \frac{-\left( \lambda -2J_c\Delta _c\cos
k\right) \cos k}{\sqrt{\left( \lambda -2J_c\Delta _c\cos k\right)
^2+(2J_s\Delta _s\sin k)^2}} &=&\Delta _c,  \nonumber \\
\int_{-\pi }^\pi \frac{dk}\pi \frac{2J_s\sin ^2k}{\sqrt{\left( \lambda
-2J_c\Delta _c\cos k\right) ^2+(2J_s\Delta _s\sin k)^2}} &=&1,  \nonumber \\
\int_{-\pi }^\pi \frac{dk}\pi \frac{\left( \lambda -2J_c\Delta _c\cos
k\right) }{\sqrt{\left( \lambda -2J_c\Delta _c\cos k\right) ^2+(2J_s\Delta
_s\sin k)^2}} &=&1.
\end{eqnarray}
In particular, when $V/J=4$, the self-consistent equations are easily
solved, and we obtain $\lambda =4J/\pi $, $\Delta _c=2\sqrt{2}/\pi $, and $%
\Delta _s=0$. There are four degenerate gapless fermionic energy bands,
different from the three bosonic elementary excitations obtained from the
Bethe ansatz method \cite{sutherland}. However, as will be shown later,
there are only three gapless collective (bosonic) excitations, so the
physical conclusions are correct. The reason why four, instead of three,
gapless modes show up in the fermion representation is similar to the weak
coupling limit of the effective bosonization approach\cite{azaria}. In that
approach the charge excitation becomes gapped in the strong coupling limit
due to an umklapp term, whereas the other three branches remain degenerate
and gapless with marginally irrelevant interactions \cite{azaria}. Away from
the SU(4) symmetric point $V/J<4$, numerical calculations can be performed
and solutions to these self-consistent equations are derived: $\lambda $ and 
$\Delta _c$ steadily decrease as the coupling parameter $V/J$ is reduced,
while $\Delta _s$ gradually increases at the same time.

The ground state energy per site is plotted as a function of the coupling
parameter $V/J$ in Fig.1. In order to compare two limiting cases, the
corresponding ground state energies for $\Delta _c\neq 0$, $\Delta _s=0$ and 
$\Delta _c=0$, $\Delta _s\neq 0$ are also plotted in the same figure as
well. It has been found that the gapped phase ($\Delta _c\neq 0$ and $\Delta
_s\neq 0$) is smoothly connected with the strong coupling SU(4) symmetric
gapless phase ($\Delta _c\neq 0,\Delta _s=0$), and represents the possible
lowest ground energy state in the parameter range of $0.462<V/J\leq 4$. Our
strong coupling MF theory is probably limited to this regime.

In Fig.2 the fermionic quasiparticle spectra with both positive and negative
energies are plotted in the range $\left[ -\pi /2,\pi /2\right] $\ for
different couplings $V/J=3.85$, $2.00$, $1.265$, $0.8$. In particular, at a
special value of $(V/J)_c\sim 1.265$, both spectra in the range $\left[
-k_F,k_F\right] $ become completely flat. Furthermore, for $V/J>(V/J)_c$,
the minimum point of the spectra appears at a finite Fermi momentum $\pm k_F$%
, while for $V/J<$ $(V/J)_c$\ the minimum point of the spectra is shifted to
zero. The special value of $(V/J)_c$ is very close to the exactly soluble
point of $V/J=4/3$ of the model Hamiltonian \cite{km}.

The Fermi momentum in the excitation spectrum mentioned above is given by 
\begin{equation}
k_F=\cos ^{-1}\left[ \frac{\lambda (2J_c\Delta _c)}{(2J_c\Delta
_c)^2-(2J_s\Delta _s)^2}\right] ,
\end{equation}
which is also calculated as a function of $V/J$ and plotted in Fig.3a. It
has been found that $k_F$ is almost fixed at $\pi /4$ over a large range of $%
2\leq V/J\leq 4$, and then quickly decreases to zero near the point of $%
V/J\sim 1.18$. The energy gap opens up at momentum $k_F$ and is evaluated as 
\begin{equation}
\Delta _{gap}=\sqrt{(2J_s\Delta _s)^2+\frac{\lambda ^2(2J_s\Delta _s)^2}{%
|(2J_c\Delta _c)^2-(2J_s\Delta _s)^2|}},
\end{equation}
which is presented in Fig. 3b. In the range of $3<V/J<4$, the energy gap is
extremely small. This agrees with the recent density matrix renormalization
group calculations showing exponentially slow gap opening \cite{ysu}, in
contrast to the earlier results \cite{pati}. Only when $V/J<3$, the energy
gap starts to grow slowly up to a maximum near the critical value $%
(V/J)_c\sim 1.265$, and then decreases to a very small value near $V/J\sim
0.462$. The position of the maximum energy gap roughly corresponds to the
condition of the dispersionless quasiparticle excitations, which is
consistent with the analysis at the exactly soluble point \cite{km}.

In the present strong coupling MF theory, the spin-spin and orbital-orbital
density correlation functions can simply be evaluated as well. The spin and
orbital density operators Eq. (4) are re-expressed in terms of the
generalized Nambu spinor 
\begin{eqnarray*}
S_i^\alpha &=&\frac 14\Psi _i^{\dagger }\Omega _S^\alpha \Psi _i,\text{ }%
T_i^\alpha =\frac 14\Psi _i^{\dagger }\Omega _T^\alpha \Psi _i, \\
\Omega _S^x &=&\sigma _z\otimes \sigma _0\otimes \sigma _x,\text{ }\Omega
_S^y=\sigma _0\otimes \sigma _0\otimes \sigma _y, \\
\Omega _S^z &=&\sigma _z\otimes \sigma _0\otimes \sigma _z,\text{ }\Omega
_T^x=\sigma _z\otimes \sigma _x\otimes \sigma _0, \\
\Omega _T^y &=&\sigma _0\otimes \sigma _y\otimes \sigma _0,\text{ }\Omega
_T^z=\sigma _z\otimes \sigma _z\otimes \sigma _0.
\end{eqnarray*}
Then the spin and orbital density-density correlation functions are given by 
\begin{equation}
\chi _X^\alpha (q,i\omega _m)=-\frac 1{16\beta }\sum_{\omega _n}\int \frac{dk%
}{2\pi }{\rm Tr}\left[ \Omega _X^\alpha {\bf G}(k,i\omega _n)\Omega
_X^\alpha {\bf G}(k+q,i\omega _m+i\omega _n)\right] ,
\end{equation}
where $\Omega _X^\alpha =\Omega _S^\alpha $ for the spin and $\Omega
_X^\alpha =\Omega _T^\alpha $ for the orbital. By inserting the Matsubara
Green function, it is straightforward to prove the following relation, 
\begin{equation}
\chi _S^\alpha (q,i\omega _m)=\chi _T^\alpha (q,i\omega _m)\equiv \chi
(q,i\omega _m),
\end{equation}
independent of the indices $\alpha =x,y,z$. This shows that away from the
SU(4) symmetry point, the spin and orbital rotational symmetry of SU(2)$%
\otimes $SU(2) with an additional Z$_2$ symmetry in exchange between ${\bf S}%
_i$ and ${\bf T}_i$ is satisfied in the present strong coupling MF state.
The resulting expression of $\chi (q,i\omega _m)$ is given by 
\begin{eqnarray}
\chi (q,i\omega _m) &=&-\frac 1{2\beta }\sum_{\omega _n}\int \frac{dk}{2\pi }
\nonumber \\
&&\frac{i\omega _n(i\omega _m+i\omega _n)+\left( \lambda -\Delta
_c(k)\right) \left( \lambda -\Delta _c(k+q)\right) +\Delta _s(k)\Delta
_s(k+q)}{\left[ \left( i\omega _n\right) ^2-\epsilon _k^2\right] \left[
\left( i\omega _m+i\omega _n\right) ^2-\epsilon _{k+q}^2\right] }.
\end{eqnarray}

However, in the SU(4) spin-fermion representation, nine spin-orbital tensor
operators associated with nonlinear collective excitation of both spin and
orbital degrees of freedom, can be defined by $L_i^{\alpha ,\beta
}=2S_i^\alpha T_i^\beta $, which can also be expressed in terms of the
generalized Nambu spinor as 
\begin{eqnarray}
L_i^{\alpha ,\beta } &=&\frac 14\Psi _i^{\dagger }\Omega _L^{\alpha ,\beta
}\Psi _i,  \nonumber \\
\Omega _L^{xx} &=&\sigma _z\otimes \sigma _x\otimes \sigma _x,\text{ }\Omega
_L^{xy}=\sigma _0\otimes \sigma _y\otimes \sigma _x,  \nonumber \\
\Omega _L^{xz} &=&\sigma _z\otimes \sigma _z\otimes \sigma _x,\text{ }\Omega
_L^{yx}=\sigma _0\otimes \sigma _x\otimes \sigma _y,  \nonumber \\
\Omega _L^{yy} &=&\sigma _z\otimes \sigma _y\otimes \sigma _y,\text{ }\Omega
_L^{yz}=\sigma _0\otimes \sigma _z\otimes \sigma _y,  \nonumber \\
\Omega _L^{zx} &=&\sigma _z\otimes \sigma _x\otimes \sigma _z,\text{ }\Omega
_L^{zy}=\sigma _0\otimes \sigma _y\otimes \sigma _z,  \nonumber \\
\text{ }\Omega _L^{zz} &=&\sigma _z\otimes \sigma _z\otimes \sigma _z.
\end{eqnarray}
Then it can be proved that the corresponding nine correlation functions $%
\langle {\rm T}_\tau L_i^{\alpha ,\beta }(\tau )L_j^{\alpha ,\beta }(\tau
^{\prime })\rangle $ are the same, and equal to 
\begin{eqnarray}
\chi _L(q,i\omega _m) &=&-\frac 1{2\beta }\sum_{\omega _n}\int \frac{dk}{%
2\pi }  \nonumber \\
&&\frac{i\omega _n(i\omega _m+i\omega _n)+\left( \lambda -\Delta
_c(k)\right) \left( \lambda -\Delta _c(k+q)\right) -\Delta _s(k)\Delta
_s(k+q)}{\left[ \left( i\omega _n\right) ^2-\epsilon _k^2\right] \left[
\left( i\omega _m+i\omega _n\right) ^2-\epsilon _{k+q}^2\right] }.
\end{eqnarray}
Compared to the spin and orbital density correlation functions, there is
only a sign difference in front of $\Delta _s(k)\Delta _s(k+q)$. In the
conventional response function theory, the correlation spectrum $\chi
_L(q,i\omega _m)$ represents a {\it nonlinear} collective excitations of
both spins and orbitals. When $V/J=4$, it is found that $\Delta _s=0$, and
then we have $\chi _L^{\alpha \beta }(q,i\omega _m)=\chi _S^\alpha
(q,i\omega _m)=\chi _T^\alpha (q,i\omega _m)$, independent of their
component indices of spin and orbital operators, which implies that the
SU(4) symmetry is recovered.

After summation over the Matsubara frequency and analytical continuation, we
find that in both dynamic susceptibilities $\chi (q,\omega )$ and $\chi
_L(q,\omega )$, an energy gap exists at $\omega =2\Delta _{gap}$ in the
large parameter range of $0.462<V/J<4$. As far as the relevant experiments 
\cite{axtell,isobe} are concerned, however, the most important physical
quantities measured are the corresponding static susceptibilities. In the
zero frequency limit, the static susceptibilities $\chi (q)$ and $\chi _L(q)$
are deduced to respectively 
\begin{eqnarray*}
\left( 
\begin{array}{l}
\chi (q) \\ 
\chi _L(q)
\end{array}
\right) &=&\int \frac{dk}{8\pi }\left\{ \frac{\left[ n_F(\epsilon
_k)-n_F(\epsilon _{k+q})\right] }{\epsilon _{k+q}-\epsilon _k}\left[ 1+\frac{%
\left( \lambda -\Delta _c(k)\right) \left( \lambda -\Delta _c(k+q)\right)
\pm \Delta _s(k)\Delta _s(k+q)}{\epsilon _k\epsilon _{k+q}}\right] \right. \\
&&+\left. \frac{\left[ 1-n_F(\epsilon _{k+q})-n_F(\epsilon _k)\right] }{%
\epsilon _{k+q}+\epsilon _k}\left[ 1-\frac{\left( \lambda -\Delta
_c(k)\right) \left( \lambda -\Delta _c(k+q)\right) \pm \Delta _s(k)\Delta
_s(k+q)}{\epsilon _k\epsilon _{k+q}}\right] \right\} .
\end{eqnarray*}
We present the static spin and orbital susceptibility at zero temperature in
Fig.4 and the static spin-orbital tensor susceptibility at zero temperature
in Fig.5 for different coupling parameters $V/J=2.72$, $2.00$, $1.265$, $%
0.80 $, and $0.50$.

At $V/J=2.72$, since the energy gap in the elementary excitations is
negligible, both static susceptibilities $\chi (q)$ and $\chi _L(q)$ have
sharp peaks at $q=\pm \pi /2$, corresponding to a commensurate spin- and
orbital-density wave of period four. The period four arises because the spin
and orbital chain are in their SU(4) fundamental representation and by
''quadruplicity'' one needs four sites to form a singlet. Moreover, as the
momentum goes to zero, $\chi (q)$ approaches to zero, while $\chi _L(q)$ to
a constant.

As $V/J$ decreases further, a finite gap opens up, and the sharp peaks in
both $\chi (q)$ and $\chi _L(q)$ spectra at $q=\pm \pi /2$ are strongly
suppressed and broadened. Meanwhile, the peak positions are slightly shifted
to lower momenta, indicating the possible presence of incommensurate density
waves in the gapped phase. The spectral weights of the coherent
quasiparticle peaks are gradually transferred to the incoherent excitations
at $q=\pm \pi $ for $\chi (q)$ and at $q=0$ for $\chi _L(q)$. Thus, the
static spin and orbital susceptibilities around $q=\pm \pi $ are enhanced,
exhibiting a crossover from coherent to incoherent magnetic excitations.
Some of these features have been speculated by an effective low-energy
bosonization theory \cite{azaria,tsvelik}. From our MF theory, the crossover
is estimated to occur around the critical coupling $V/J=1.265$. It seems to
us that the present symmetric state is a good approximation of the genuine
ground state to describe this crossover. Moreover, the spectrum of $\chi
_L(q)$ in Fig.5 displays a further clear evidence of the crossover of the
magnetic excitations from the strong to weak coupling limits of the system.
In particular, when $V/J$ becomes smaller and smaller, a new peak structure
clearly emerges at zero momentum. All these features of $\chi _L(q)$ are new
results obtained from the present non-perturbative theory.

Finally, several remarks and comments are in order. i) The fermionic MF
theory developed in this paper mainly focuses on the energy gap formation of
the spin-orbital coupled model, in particular, on the crossover of
coherent-incoherent magnetic excitations in the gapped phase. ii) All the
obtained results on the gapped phase are mostly consistent with the latest
density matrix renormalization group calculation \cite{ysu} and the
effective bosonization theory \cite{azaria}. One of the important results is
that the energy gap opens up extremely slowly away from the SU(4) symmetric
point, indicating that the phase transition from gapless to gapped phases
may be of Kosterlitz-Thouless type. The gapped regime with coherent magnetic
excitations shares some similarities with the Haldane gapped phase of the
quantum antiferromagnetic spin-ladder model with a four-spin interaction 
\cite{tsvelik}. So it is not clear whether the Lieb-Schultz-Mattis theorem
is valid in the present spin-orbital coupled system or not. iii) The gapless
phase at the SU(4) point is only characterized by a Fermi liquid behavior in
the present MF theory approximately, different from the Luttinger liquid
behavior derived from the Bethe ansatz and effective field theories \cite
{sutherland,affleck,azaria}. However, as far as the spin, orbital, and
spin-orbital tensor collective excitations are concerned, all these physical
results of the spin-orbital system --- the correlation spectra are in good
agreement with the exact solution.

In conclusion, we have applied an SU(4) constraint fermion representation to
a one-dimensional symmetrically coupled spin-orbital model, a spin-orbital
gapped phase is generated away from the strong coupling SU(4) symmetric
point by a relevant spin-orbital pairing interaction. The energy gap of the
elementary spin and orbital excitations grows extremely slowly up to a
maximum value and then decreases. The spin, orbital, and spin-orbital tensor
static susceptibilities are also calculated at zero temperature, displaying
a crossover from coherent to incoherent magnetic excitations as the
spin-orbital coupling decreasing away from the SU(4) symmetry point. It is
interesting to note that the present mean field theory provides a rather
good description of the underlying physics over a large crossover region
between the two critical points. It's likely that the SU(4) constraint
fermion representation is appropriate for this system.

We would like to thank Prof. Z. B. Su and X. Q. Wang for useful discussions,
while L. Yu would like to thank A. Nersesyan for very helpful comments. This
work is supported by NSF-China (Grant No. 10074036 and 10125418) and the
Special Fund for Major State Basic Research Projects of China (G2000067107).

Figure Captions

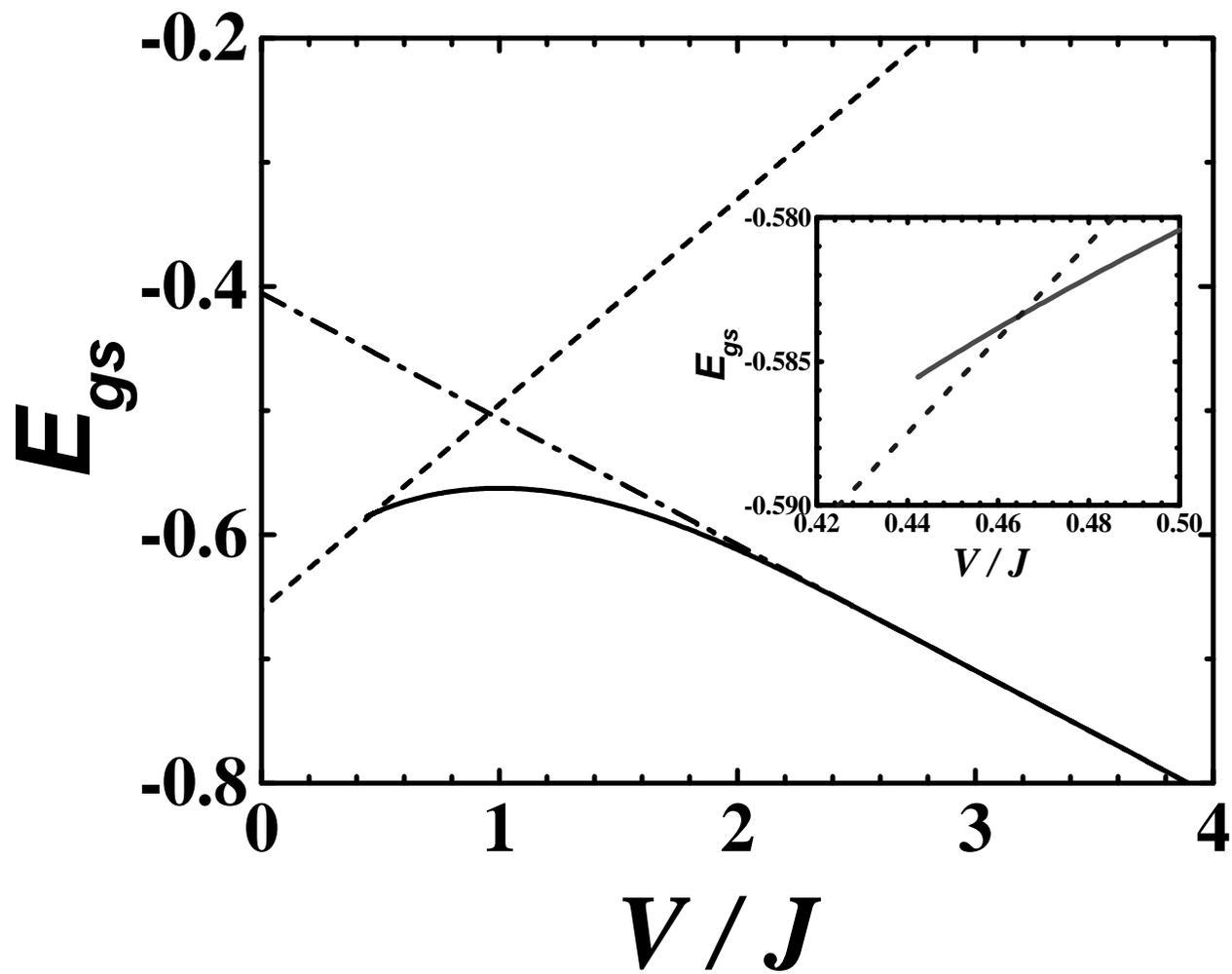
Fig. 1. The ground state energy per site as a function of the coupling
parameter $V/J$. For comparison, the corresponding ground state energies for 
$\Delta _c\neq 0,\Delta _s=0$ and $\Delta _c=0,\Delta _s\neq 0$ are also
plotted by the dashed and dash-dotted lines, respectively.

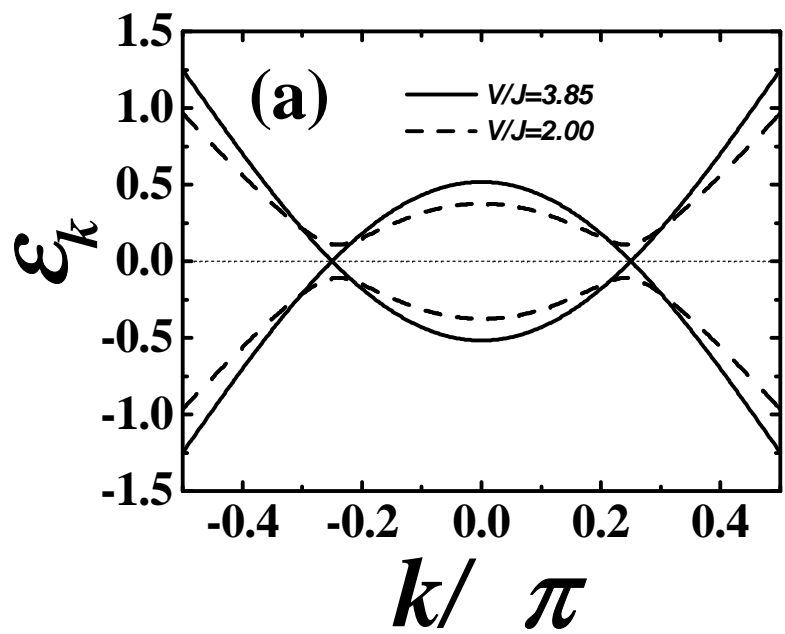
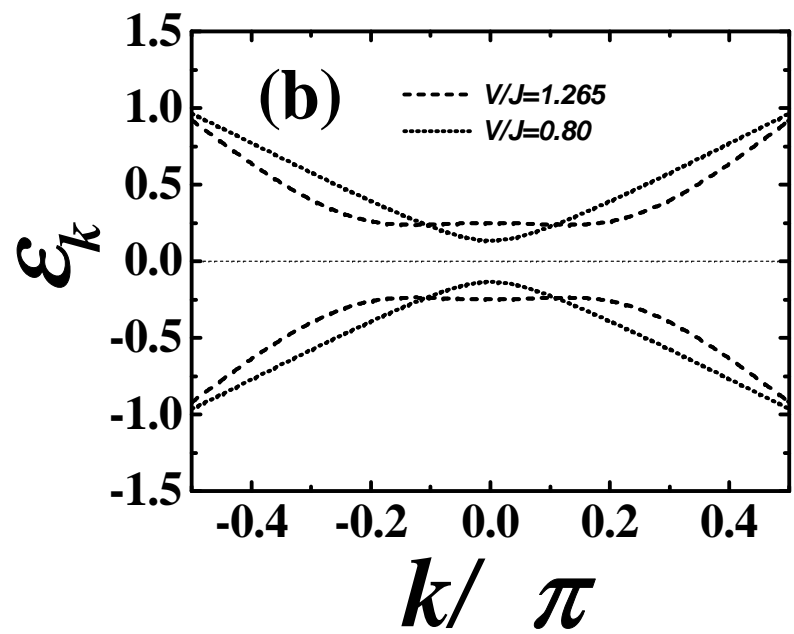
Fig. 2. The fermionic energy spectra are plotted in the range $\left[ -\pi
/2,+\pi /2\right] $ for the coupling values: a) $V/J=3.85$, $2.00$ and b) $%
V/J=1.265$, $0.80$.

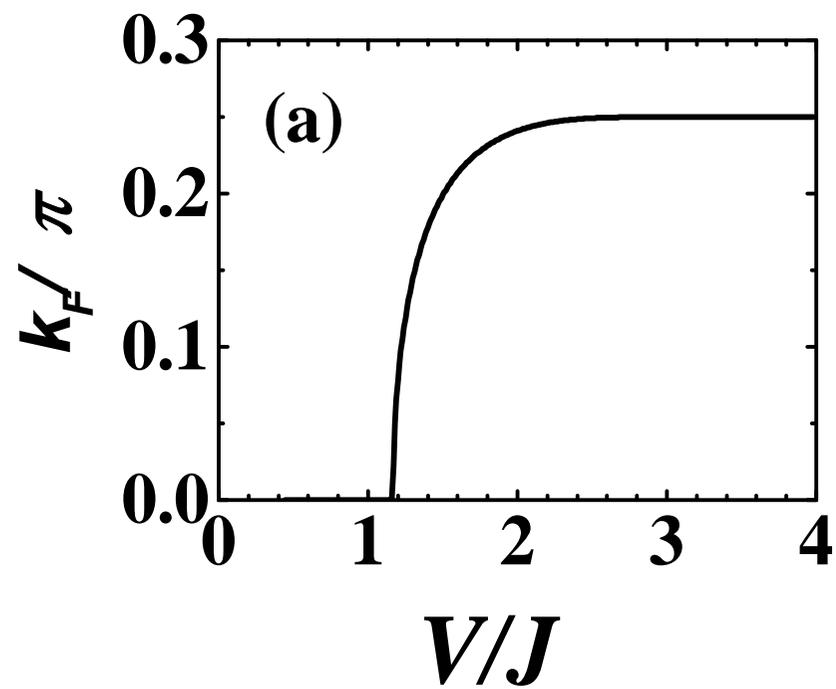
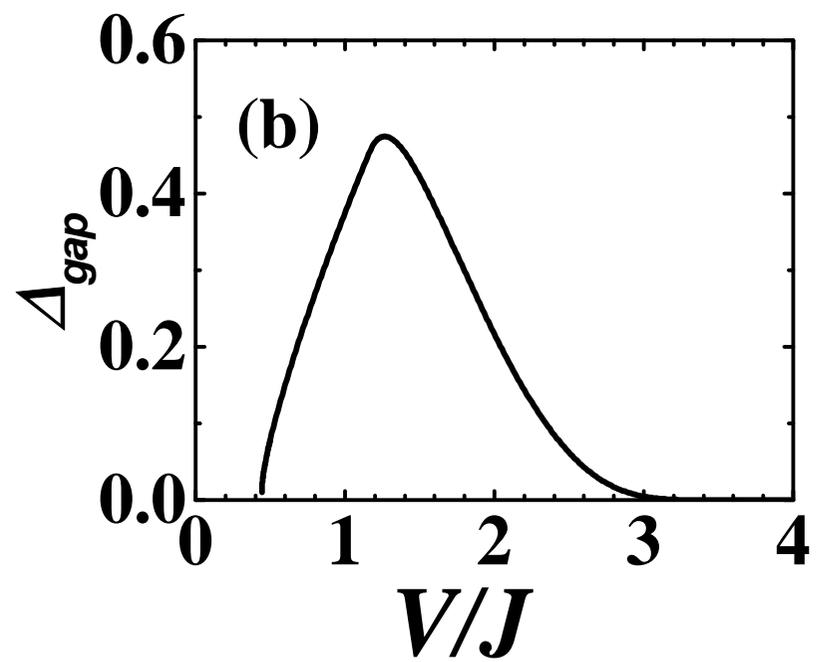
Fig. 3. The Fermi momentum $k_F$ and the corresponding quasiparticle gap as
functions of coupling parameter $V/J$.

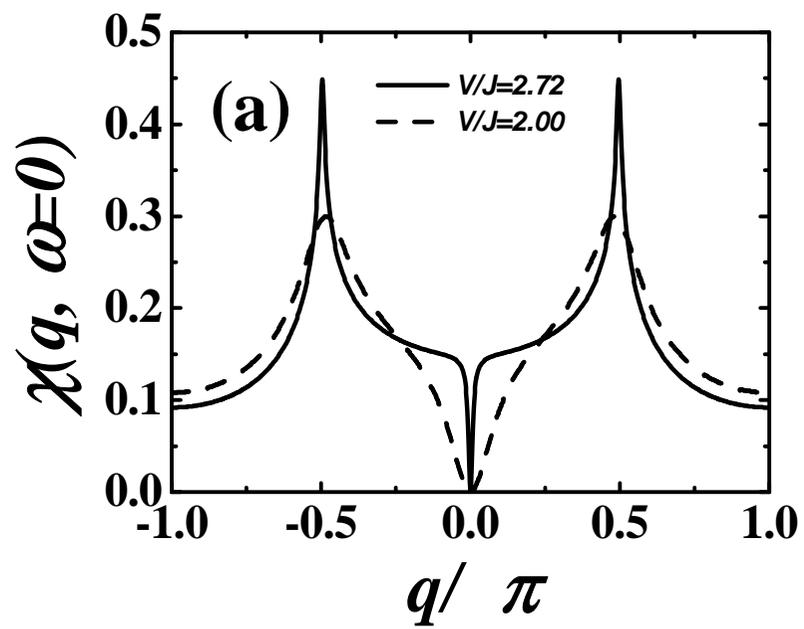
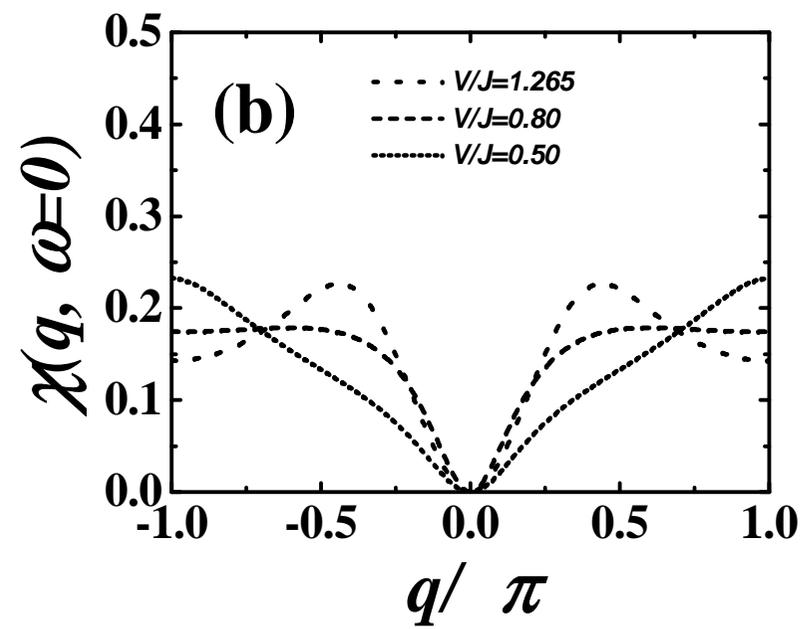
Fig. 4. The static spin and orbital susceptibilities at zero temperature for
different coupling values: a) $V/J=2.72$, $2.00$ and b) $V/J=1.265$, $0.80,$
and $0.50$.

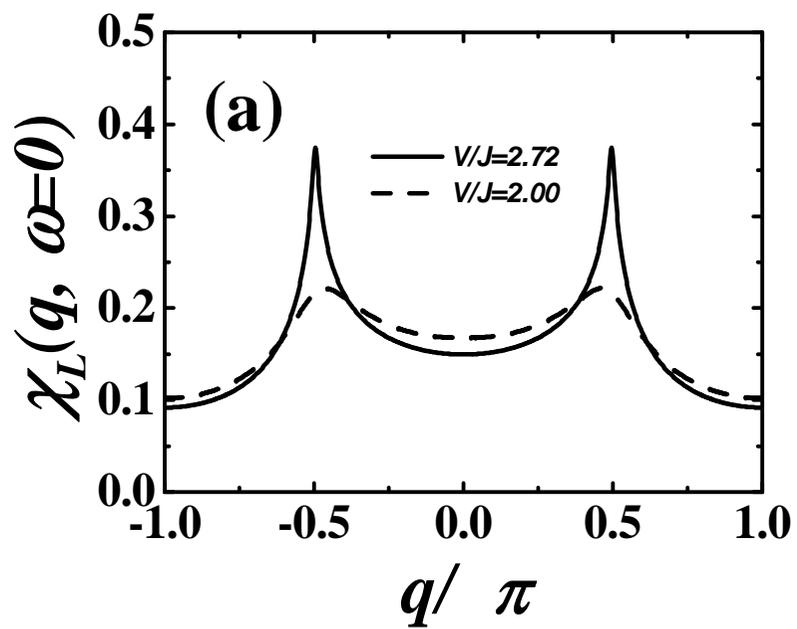
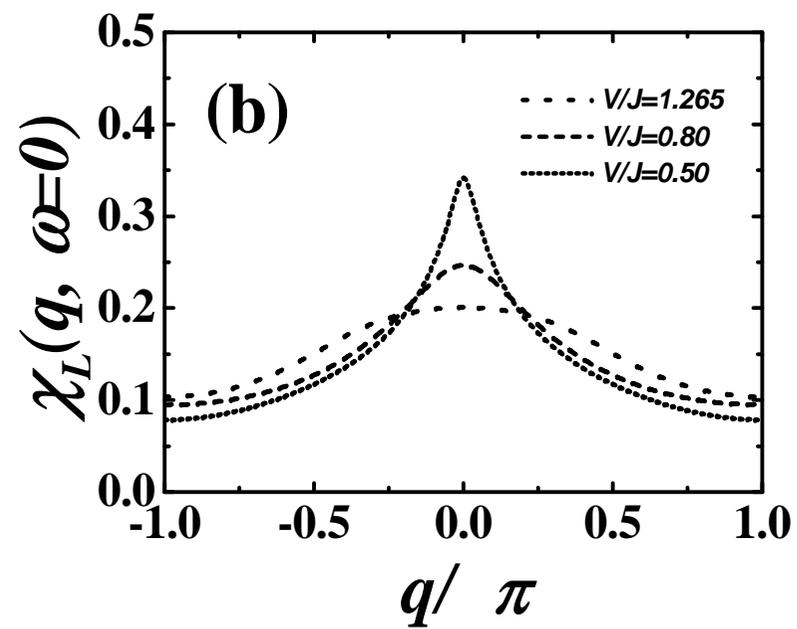
Fig. 5. The static susceptibility of the spin-orbital tensor at zero
temperature for different coupling values: a) $V/J=2.72$, $2.00$ and b) $%
V/J=1.265$, $0.80,$ and $0.50$.


\begin{references}
\bibitem{convention}  Hereafter we use wording ''weak'' and ''strong''
coupling only meaning small and large value of $V/J$, without any reference
to the renormalization group flow.
\end{references}
\end{document}